\documentclass[conference]{IEEEtran}

\usepackage{graphicx,cite,epsfig}
\usepackage{amsmath}
\usepackage{amssymb}
\DeclareMathOperator *{\argmin}{argmin}
\usepackage{pifont}
\usepackage{stmaryrd}
\usepackage{bbding}
\usepackage{subfigure}
\usepackage{algorithm}
\usepackage{algorithmic}
\usepackage{array}
\usepackage{amsthm}
\usepackage{url}
\usepackage{mathrsfs}
\usepackage{color, soul}
\usepackage{flushend}
\usepackage{mathrsfs}
\usepackage{epstopdf}

\begin{document}
\title{Adaptive Perturbation Enhanced SCL Decoder for Polar Codes}
\author{\IEEEauthorblockN{Xianbin~Wang, Huazi~Zhang, Jiajie~Tong, Jun~Wang, Wen~Tong}
	\IEEEauthorblockA{Huawei Technologies Co., Ltd.\\
	Emails: \{wangxianbin1,zhanghuazi,tongjiajie,justin.wangjun,tongwen\}@huawei.com}}\maketitle

\maketitle
\thispagestyle{empty}

\begin{abstract}
For polar codes, successive cancellation list (SCL) decoding algorithm significantly improves finite-length performance compared to SC decoding.
SCL-flip decoding can further enhance the performance but the gain diminishes as code length increases, due to the difficulty in locating the first error bit position.
In this work, we introduce an SCL-perturbation decoding algorithm to address this issue.
A basic version of the algorithm introduces small random perturbations to the received symbols before each SCL decoding attempt, and exhibits non-diminishing gain at large block lengths.
Its enhanced version adaptively performs random perturbations or directional perturbation on each received symbol according to previous decoding results, and managed to correct more errors with fewer decoding attempts.
Extensive simulation results demonstrate stable gains across various code rates, lengths and list sizes.
To the best of our knowledge, this is the first SCL enhancement with non-diminishing gains as code length increases, and achieves unprecedented efficiency. With only one additional SCL-$L$ decoding attempt (in total two), the proposed algorithm achieves SCL-$2L$-equivalent performance.
Since the gain is obtained without increasing list size, the algorithm is best suited for hardware implementation\footnote{Part of this paper was presented at the 2023 IEEE Global Communications Conference~\cite{GC2023}.}.
\end{abstract}


\section{Introduction}\label{introduction}
\subsection{Polar codes}
Polar codes are the first capacity-achieving channel codes with low-complexity successive cancellation (SC) decoding~\cite{ArikanTIT09}. However, their finite-length performance was considered poor until the introduction of successive cancellation list (SCL) and CRC-aided SCL (CA-SCL) decoding algorithms~\cite{TalVardy,KaiNiu}. With these enhanced decoding algorithms, polar codes outperform Turbo codes and LDPC codes at short to medium block lengths, and thus are very attractive for encoding short messages in practical communication systems. In 2017, polar codes were adopted as the channel coding scheme for control channel in the 5G NR standard~\cite{3GPP}.

From the practical viewpoint, decoding algorithms are the most important factor when evaluating the feasibility of implementing a channel coding scheme. For polar codes, several decoding algorithms have been proposed in literature, and continuous efforts have been made to significantly enhance performance and reduce decoding complexity~\cite{flip,DynamicFlip,Shifted,Access,Viterbo1,Viterbo2,Xiaohu,Burg1,Krouk1,Mingmin,PC,AEBrink,AEBioglio,AElistBrink}.

SC decoder has very low complexity, thus is suitable for applications requiring extremely high throughput~\cite{Fast} or low power~\cite{LowPower}. But its error correction performance is mediocre. SCL decoding is more complex than an SC decoder because it maintains a list of SC decoder instances in parallel, and requires list management to keep the good codeword candidates during decoding. In practice, the list size $L$ is limited to a small number (e.g., $L=8\sim32$) to strike a good balance between performance and complexity.

\subsection{Enhancements to SC/SCL decoding}
For SCL decoding, the main hardware constraint is the list size. Motivated by this, researchers explored alternative methods to improve decoding performance without increasing the list size. In the following, we review two major enhancements.

A low-complexity yet efficient approach is SC/SCL-flip decoding~\cite{flip,DynamicFlip,Shifted,Access,Viterbo1,Viterbo2,Xiaohu,Burg1,Krouk1,Mingmin,PC}. Instead of increasing list size, the algorithm develops a small number of list paths at a time. In case of a failed decoding attempt (detected by CRC check), a new attempt is made to identify and flip the first bit errors encountered during the previous attempts. Subsequent decoding attempts are made until a correct codeword is found (e.g., passes CRC check), or a maximum number of attempts is reached.
\begin{figure}[tbp]
	\centering
	\includegraphics[width=2.8in]{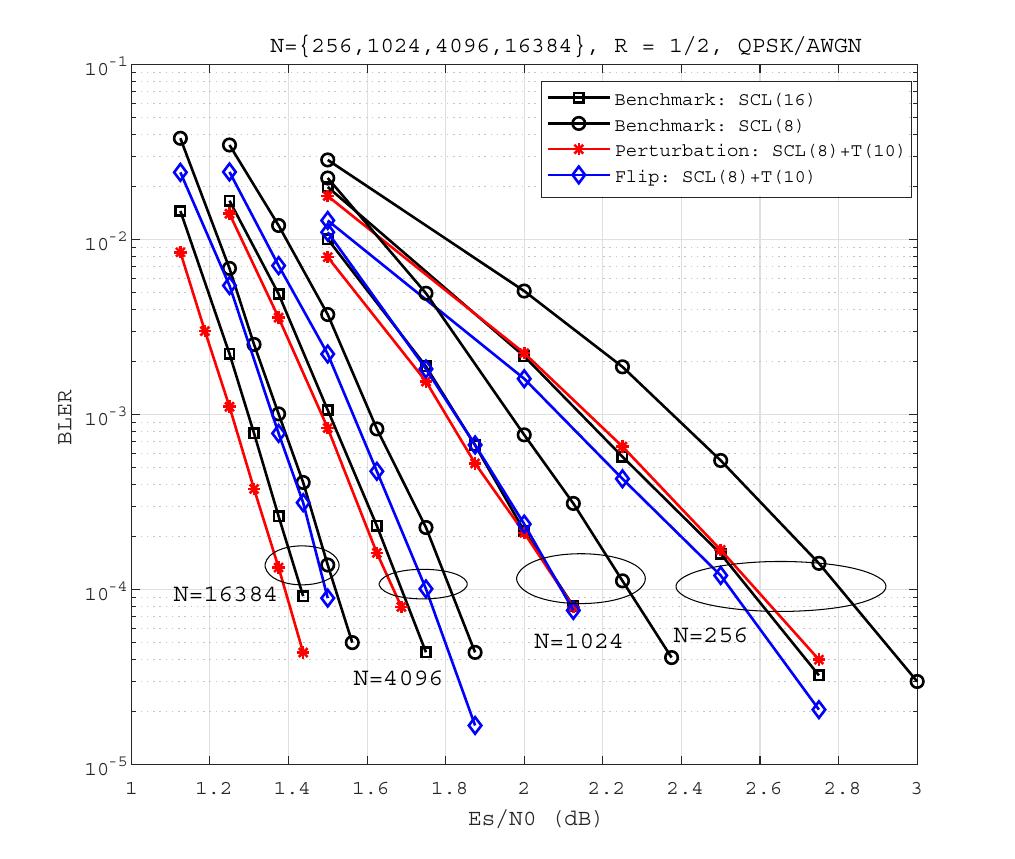}
	\caption{BLER performance of SCL($8\sim16$) decoding, SCL perturbation with T = 10 decoding attempts and SCL flip with T = 10 for polar codes of $R=0.5$ and $N=\{256,1024,4096,16384\}$.}
	\label{simualtion}
\end{figure}

Th flipping approach has been extensively studied in the SC decoding (see~\cite{DynamicFlip} and the references therein for more details). The technique was later adapted to SCL decoding~\cite{Shifted,Access,Viterbo1,Viterbo2,Xiaohu,Burg1,Krouk1,Mingmin,PC}. Although SC/SCL-flip decoding can further improve the SC/SCL performance, they have two main shortcomings. First, the gain quickly vanishes as code length increases~\cite{Access}, because it becomes more difficult to locate the first error bit from a larger unreliable information bit set. Second, the worst-case decoding latency also increases with code length.

Another recently proposed approach is the automorphism ensemble (AE) decoding~\cite{AEBrink,AEBioglio,AElistBrink}. Specifically, a set of permutations are generated according to polar codes' automorphisms. The SC/SCL-based AE decoders apply these permutations to received symbols. Each permuted version of received symbols is decoded by an SC/SCL decoder, and inversely permuted to recover a codeword candidate. By properly choosing the SC-variant permutations, these codeword candidates will be different. Finally, the most likely candidate is selected as the decoding output. Automorphism ensemble provides a diversity gain similar to that of a list decoder, and thus improves decoding performance.
\begin{figure}[tbp]
	\centering
	\includegraphics[width= 3.1in]{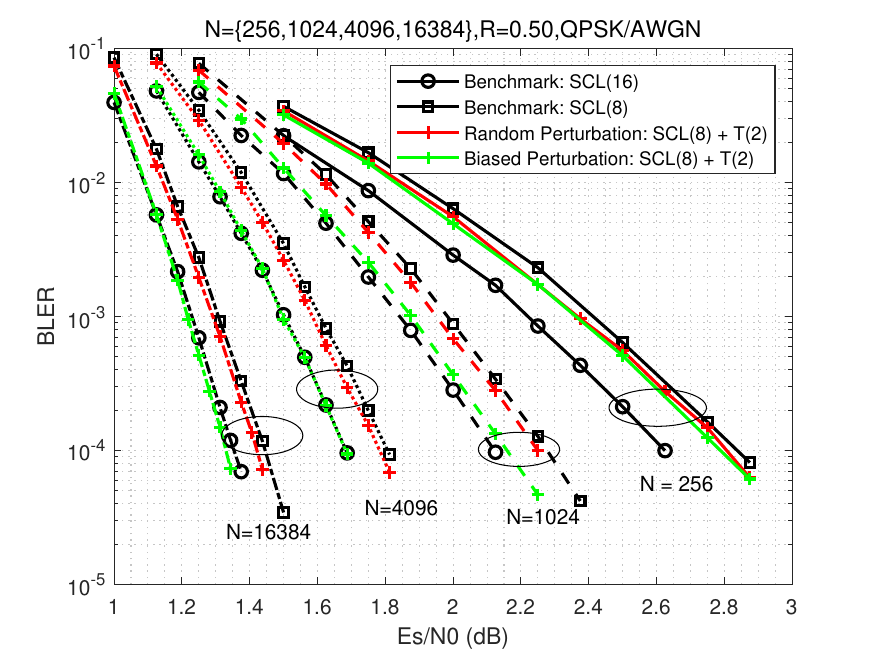}
	\caption{BLER performance of SCL ($8\sim16$) decoding, SCL-perturbation (random and bias) with T = 2 decoding attempt for polar codes of $R=0.5$ and $N=\{256,1024,4096,16384\}$.}
	\label{iSCL_performance_05}
\end{figure}

In practice, SC/SCL-based AE decoders face several challenges. Exsiting research efforts mainly include finding sufficient SC-variant permutations~\cite{YuanLiAE} to promote decoding diversity. This turns out to be difficult as code length increases, because the group of affine automorphisms of polar codes cannot be significantly larger than the group of SC-invariant lower-triangular permutations in the asymptotic sense~\cite{Ivanov}. Similar to SC/SCL-flip, the benefit of SC/SCL-AE diminishes at longer block lengths.

\subsection{Contributions}
In this paper, we propose an SCL-perturbation decoding algorithm to address this issue. This algorithm also leverages multiple SCL decoding attempts. However, we add small random perturbations to the received symbols before each SCL decoding attempt. By properly setting the perturbation power, the decoder is able to collect a larger set of highly-likelihood codewords, mimicking the effect of a larger list size, and thus increases the possibility of successful decoding.

This algorithm provides a stable and non-diminishing gain as code length increases (see Fig.~\ref{simualtion}).
With the same number of decoding attempts, the \emph{perturbation gain} is higher than the \emph{flipping gain} at larger code lengths. Specifically, with a maximum of $9$ perturbed SCL-$L$ decoding attempts ($10$ in total including the initial decoding), the algorithm can achieve the performance of an SCL-$2L$ decoder.

To achieve the same performance with fewer decoding attempts, we introduce a ``biased SCL-perturbation'' method to combine random perturbation and directional perturbation. The directional perturbation generates constant perturbation values based on previous decoding outcomes. Although the decoding results from prior attempts are erroneous, they still contain useful information. For instance, when all previous decoding results suggest that a code bit is positive, the true value of this bit is more likely to be positive. By directional perturbation, a received symbol can be pulled toward its true value.

This modification turns out to be surprisingly effective, i.e., achieving the same gain with fewer decoding attempts. As shown in Fig.~\ref{iSCL_performance_05}, biased SCL-perturbation outperforms random SCL-perturbation in the low-complexity regime. With a maximum of one extra SCL-$L$ decoding attempt (two including the initial decoding), it achieves SCL-$2L$ performance.

The contributions of this paper are summarized as follows:
\begin{enumerate}
	\item We propose a random SCL-perturbation decoding algorithm, which, to the best of the our knowledge, is the first SCL-enhancement with non-diminishing gains as code lengths increase. Extensive simulation results under various code rates, code lengths, list sizes, and numbers of decoding attempts confirm the efficiency of the proposed algorithm.
	
	\item We propose a biased SCL-perturbation decoding algorithm to further improve decoding efficiency. Specifically, we apply directional perturbations to some highly-confident code bit positions, along with random perturbations on the remaining bit positions. This enhanced algorithm requires much fewer decoding attempts to achieve the same performance gain.
	
	\item We evaluate the decoding complexity from three perspectives: worst-case computation, average computation, and hardware implementation. To achieve the same performance, SCL-perturbation requires lower complexity than both SCL-flip and SCL with a larger list size at all signal-to-noise ratios (SNRs) of interest.
	
\end{enumerate}

Some preliminary studies have tried to enhance SC decoding with perturbation ~\cite{PerturbSC,PerturbSCDynamic,PerturbCNN,PerturbGenetic}. However, there is not a thorough research on perturbation-based SCL enhancement, including algorithm optimization, performance and complexity evaluations.

\subsection{Organization}
The remaining of this paper is organized as follows.
Section \ref{sec:preliminaries} describes the preliminaries of polar codes, including an overview of polar encoding and SC-based decoding algorithms.
Section \ref{sec:perturbation} presents the proposed SCL-perturbation algorithm with details about the random perturbation and biased perturbation methods.
In Section \ref{sec:evaluation}, extensive simulation results are provided to confirm the efficiency of the proposed algorithms.
In Section \ref{sec:complexity}, we assess the decoding complexity of the proposed algorithms. 
Section \ref{sec:discussion} discusses potential enhancements to the proposed algorithms using machine learning techniques.
Finally, Section \ref{sec:conclusion} concludes this paper.

\section{Preliminaries}\label{sec:preliminaries}
\subsection{Polar codes}
Polar Codes of length $N=2^n$ are constructed based on encoding matrix $G_N$. $G_N$ is $n$-th Kronecker power of the kernel $G_2 = \begin{bmatrix} 1 & 0 \\ 1 & 1 \end{bmatrix}$, denoted by $G_N = G_2^{\otimes n}$. The kernel transforms two equal channels to a less reliable one and a more reliable one. Channel polarization is achieved through recursively applying the kernel to produce a set of very reliable channels and a set of very noisy channels.

To construct a polar code, we select the set of reliable channels for transmitting the information bits, and do not transmit any information on the noisy channels. The encoding and modulation steps are as follows.
\begin{enumerate}
  \item An information vector $\mathbf{u}_0^{N}$ is generated by placing the $K$ information bits to the $K$ indices denoted by $\mathcal{I}$, while setting the remaining values to zero. $\mathcal{I}$ is the $K$ most reliable channel indices in $[0,\cdots,N-1]$.
  \item Multiply the information vector $\mathbf{u}_0^{N}$ by the polar matrix $G_N$ to obtain the codeword vector $\mathbf{c}_0^{N}= \mathbf{u}_0^{N}G_N$.
\end{enumerate}

The code bits are modulated by binary phase shift keying (BPSK), i.e., $x_i=1-2c_i$, and then transmitted through Gaussian channels. The received channel output is denoted by $y_i = x_i + n_i$, where $n_i\sim\mathcal{N}(0,\sigma^2)$ and $\sigma^2$ is the noise power.

\subsection{LLR-based SC/SCL decoding}
To perform decoding, each received symbol $y_i$ is first converted to a log-likelihood ratio (LLR) $L(x_i)$ according to $L(x_i) = \frac{2y_i}{\sigma^2}$.

For a length-2 polar code, SC decoding comprises one $f$-function, one $g$-function and two hard decisions, as follows.

The $f$-function calculates the LLR of bit $u_0$ based on the LLRs of $x_0$ and $x_1$ as follows:
\begin{equation}
\label{f_func}
L(u_0) = \text{sign}(L(x_0)L(x_1)) \cdot \min(|L(x_0)|,|L(x_1)|)
\end{equation}
A hard decision is made based on $L(u_0)$ to obtain $\hat{u}_0$, i.e., the estimated value of $u_0$.

Then, $g$-function is used to update the LLR of $u_1$ as follows:
\begin{equation}
\label{g_func}
L(u_1) = (1-\hat{u}_0)L(x_0)+L(x_1)
\end{equation}
Finally, another hard decision is made based on $L(u_1)$ to obtain $\hat{u}_1$.

For longer polar codes, SC decoding is performed by recursively applying the $f$- and $g$-functions, until obtaining the LLR of each information bit for hard decision. SCL decoding is built on top of SC decoding, where a list of decoding paths are developed using the SC decoder, and the $L$ most likely paths are kept after each path split at an information bit position~\cite{LLRSCL}.

\section{SCL-perturbation decoding algorithm}\label{sec:perturbation}

\begin{algorithm}[tbp]
	\caption{SCL-perturbation decoding framework}
	\label{algo:iSCL_framework}
	\begin{algorithmic}[1]
		\REQUIRE ~~\\ Received symbols $\mathbf{y}_0^N$, perturbation noise power $\sigma_p^2$, the list size $L$, the allowed decoding attempts $T$.
		\STATE $\hat{\mathbf{u}}_0^k, \hat{\mathbf{c}}_0^N$ = \text{SCLDecoding}($\mathbf{y}_0^N$, $L$) \label{line:init}
		\STATE The decoding paths set $\mathcal{P}$ is initialized with $\{\hat{\mathbf{c}}_0^N\}$.\label{line:pinit}
		\IF {\text{CRCDetection}$(\hat{\mathbf{u}}_0^k)$ = 1}\label{line:crc_check}
		\FOR{$t = 2$ to $T$} \label{line:loop_start}
		\STATE $\mathbf{p}_0^N$ = PeturbationGeneration().\label{line:perturbation_generation} 
		\STATE $\mathbf{y'}_0^N = \mathbf{y}_0^N + \mathbf{p}_0^N$.\label{line:perturb}
		\STATE $\hat{\mathbf{u}}_0^k, \hat{\mathbf{c}}_0^N$= \text{SCLDecoding}($\mathbf{y'}_0^N$, $L$) \label{line:re-decode}
		\IF {\text{CRCDetection}$(\hat{\mathbf{u}}_0^k)$ = 0} \label{line:crc_check_attempt}
		\STATE \textbf{break} \label{line:break}
		\ELSE
		\STATE $\mathcal{P} = \mathcal{P} + \{\hat{\mathbf{c}}_0^N\}$.\label{line:addnewpaths}
		\ENDIF
		\ENDFOR \label{line:loop_end}
		\ENDIF
		\STATE \textbf{Return} $\hat{\mathbf{u}}_0^k$ \label{line:return}
	\end{algorithmic}
\end{algorithm}

The SCL-perturbation decoding framework is formally described in Algorithm~\ref{algo:iSCL_framework}.
The inputs of the algorithm are the received symbols $\mathbf{y}_0^N$, the perturbation noise power $\sigma_p^2$, the list size $L$, and the allowed number of decoding attempts $T$.

We use $y'_i$ to denote the perturbed version of $y_i$:
\begin{equation}
	y'_i = y_i + p_i,
\end{equation}
in which, $p_i$ denotes the perturbation noise, the generation of which will be elaborated later.

At first, we decode the received symbols $\mathbf{y}_0^N$ using an SCL-$L$ decoder and obtain the decoding result $\hat{\mathbf{u}}_0^k$ and $\hat{\mathbf{c}}_0^N$ (Line \ref{line:init}).

We perform a CRC check on $\hat{\mathbf{u}}_0^k$. If the check passes, the algorithm returns $\hat{\mathbf{u}}_0^k$ as the final decoding output (Line \ref{line:return}).

If the CRC check fails, the algorithm enters a loop that iteratively performs the following perturb-and-decode steps (Lines~\ref{line:loop_start} to~\ref{line:loop_end}).

In each iteration, the algorithm first generates perturbation noise $\mathbf{p}_0^N$ (Line \ref{line:perturbation_generation}).
Based on this, it perturbs $\mathbf{y}_0^N$ by adding $\mathbf{p}_0^N$ to each symbol (Line \ref{line:perturb}).
Then, the algorithm repeats SCL-$L$ decoding based on $\mathbf{y'}_0^N$ to obtain new $\hat{\mathbf{u}}_0^k$ and $\hat{\mathbf{c}}_0^N$ (Line \ref{line:re-decode}).

If $\hat{\mathbf{u}}_0^k$ does not pass the CRC check, the algorithm continues the loop until $T$ attempts are reached or a successful decoding is obtained.

As the perturbation noise generation may be influenced by previous decoding outcomes, we retain all the decoding paths in the set $\mathcal{P}$ (Lines \ref{line:pinit} and \ref{line:addnewpaths}).
	
Next, we discuss the methods for generating perturbation noise.
\begin{algorithm}[tbp]
	\caption{random perturbation}
	\label{algo:randomnoise}
	\begin{algorithmic}[1]
		\REQUIRE ~~\\ Perturbation noise power $\sigma_p^2$.
		\FOR{$i = 0$ to $N-1$}
		\STATE $p_i = {n}_i$, where ${n}_i \sim \mathcal{N}(0, \sigma_p^2)$ 
		\ENDFOR
		\STATE \textbf{Return} $\mathbf{p}_0^{N}$.
	\end{algorithmic}
\end{algorithm}
	
\subsection{Random perturbation}
In this approach, we rely on random sampling to generate perturbation values, resulting in low implementation complexity.

The \textbf{random perturbation} algorithm is described in Algorithm \ref{algo:randomnoise}. It takes the perturbation noise power $\sigma_p^2$ as input.
For each symbol, it generates a perturbation value $p_i$ as $p_i = n_i$ by sampling from a normal distribution $\mathcal{N}(0, \sigma_p^2)$.
The algorithm returns the sequence of perturbation values $\mathbf{p}_0^{N}$.

The pivotal aspect of this algorithm is how to choose an appropriate perturbation power $\sigma_p^2$. Interestingly, these is a \emph{divergence-convergence tradeoff}:
\begin{itemize}
	\item Divergence to explore more codeword candidates: a higher $\sigma_p^2$ increases the distance between the perturbed symbols and the received symbols. This leads to a higher chance of obtaining new decoding paths during each decoding attempt. However, if $\sigma_p^2$ is too large, the codeword candidates will deviate far from the transmitted codeword.
	\item Convergence toward a more likely codeword candidate: a lower $\sigma_p^2$ leads to less noisy perturbed symbols, and are thus closer to the transmitted symbols. This increases the likelihood of codeword candidates obtained in each attempt, which are more likely to be the correct result. However, if $\sigma_p^2$ is too small, most the codeword candidates will be the same.
\end{itemize}

Therefore, the perturbation power $\sigma_p^2$ should not be too large or too small. The best value can be obtained by a one-dimensional search.
Specifically, we utilize the signal-to-noise ratio reduction, $\Delta$SNR, as a measure of the perturbation strength.
We observe that introducing perturbation noise within the range of $0.03$ to $0.3$ dB yields stable performance gain.

\subsection{Biased perturbation}
\label{BP}
\begin{figure*}[tbp]
	\centering
	\includegraphics[width=7.8in]{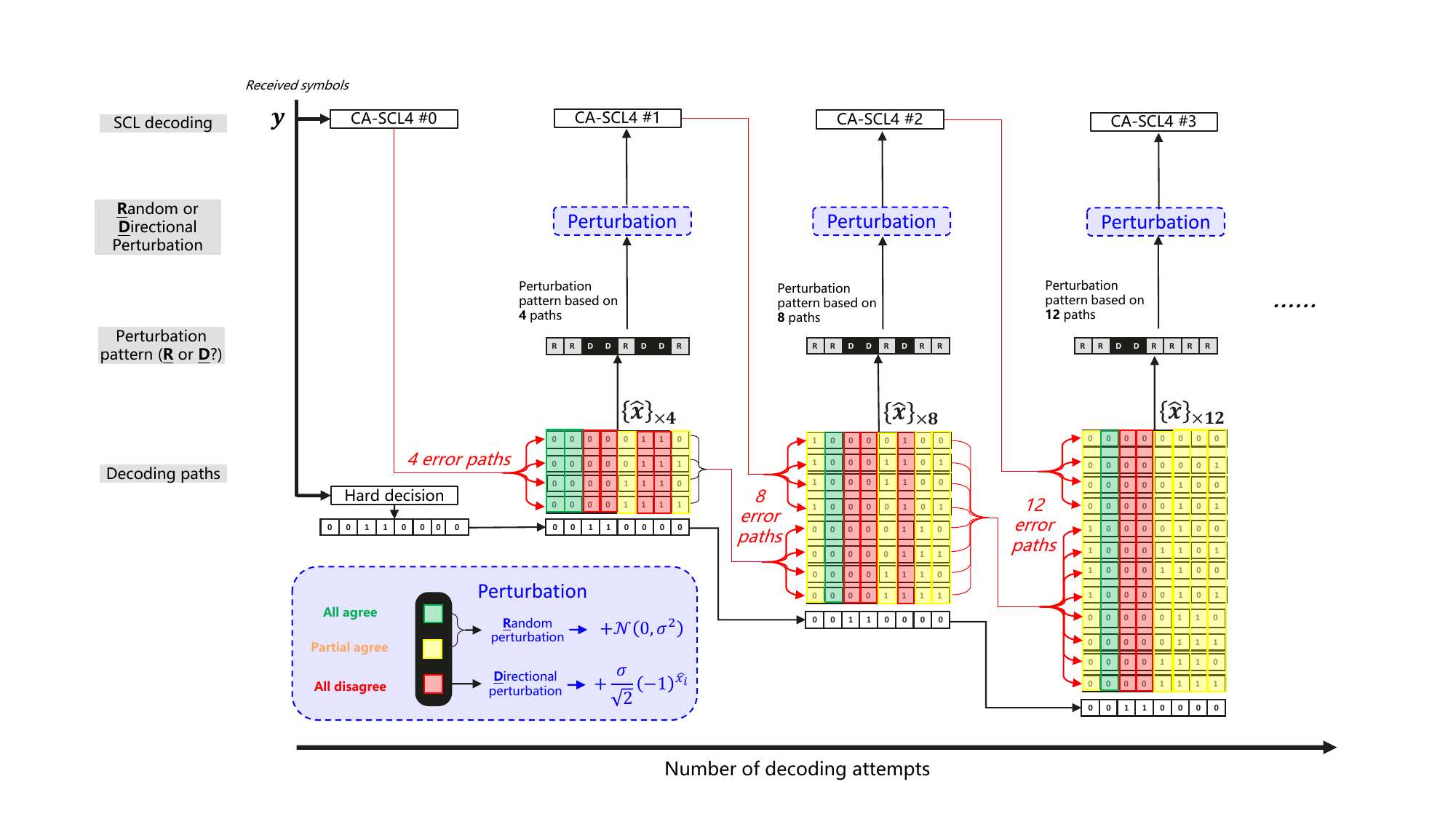}
	\caption{The algorithm applies biased perturbation only to the all-disagreed bits.}
	\label{iSCL_example}
\end{figure*}
In this approach, we determine perturbation values based on decoding results from previous decoding attempts. Although these results are erroneous from the codeword perspective, they still contain a subset of correctly decoded code bits. Our algorithm seeks to first identify these statistically more reliable bits and then properly exploit them.
For the other bits, the aforementioned random perturbation is applied, i.e., $p_i = {n}_i$, where ${n}_i \sim \mathcal{N}(0, \sigma_p^2)$.

For the identified reliable bits, a fixed and directional perturbation is applied. The perturbation values are determined as follows:
\begin{equation}
p_i = \lambda (-1)^{\hat{c_i}},
\end{equation}
where $\hat{c_i}$ denotes the decoded result of $c_i$, and $\lambda$ represents the strength of the perturbation.

\begin{algorithm}[tbp]
	\caption{biased perturbation}
	\label{algo:iterativenoise}
	\begin{algorithmic}[1]
		\REQUIRE ~~\\ Received symbols $\mathbf{y}_0^{N}$, perturbation noise power $\sigma_p^2$, the set of previous decoding paths $\mathcal{P}$.
		\FOR{$i = 0$ to $N-1$}
		\STATE $\bar{c_i} = 1_{y_i<0}$.
		\IF  {$\bar{c_i}\neq \hat{c_i}$ for all decoding paths in $\mathcal{P}$} 
		\STATE $ p_i =  \frac{\sigma_p}{\sqrt{2}}(-1)^{\hat{x_i}}$ 
		\ELSE
		\STATE $p_i = {n}_i$, where ${n}_i \sim \mathcal{N}(0, \sigma_p^2)$
		\ENDIF
		\ENDFOR
		\STATE \textbf{Return} $\mathbf{p}_0^{N}$.
	\end{algorithmic}
\end{algorithm}

Interestingly, a \emph{divergence-convergence tradeoff} is also observed in the biased perturbation process:
\begin{itemize} \item Diverging to explore more codeword candidates: Introducing random perturbations to more bits increases the distance between the perturbed symbols and the decoding results from the previous attempts. This helps to collect more new decoding results in subsequent attempts. However, the algorithm may not be able to converge toward the correct result.
	
	\item Converging toward a more likely codeword candidate: By marking more code bits as ``reliable bits'', biased perturbation may pull more of these bits towards pre-defined values. If the selected ``reliable bits'' are correct, the next decoding attempt will succeed with a higher probability. But if some of them are incorrect, the decoding results will be pulled away from the true codeword. Biased perturbation favors exploitation of previous decoding results over exploration of new decoding results. This elevates the risk of repeating the same incorrect decision in the next decoding attempt.
\end{itemize}

Therefore, a key question is how to correctly identify the subset from reliable code bits for biased perturbation. By examining the decoding results from previous decoding attempts, we find that if all previous decoding paths output the same value on certain code bits, these bits are more likely to be correct bits; but if some decoding results suggest that a code bit is positive but other results suggest it to be negative, neither results can be trusted. Based on this observation, we propose to mark the code bits with the same decoding results as reliable code bits, and those with different decoding results as unreliable code bits.

As aforementioned, there is a divergence-convergence tradeoff between exploration via random perturbation and exploitation by biased perturbation. The tradeoff is controlled by the proportion of reliable code bits selected for random or biased perturbations. To facilitate a fine-tuning of this tradeoff, we propose to further divide the reliable code bits to all-agreed bits and all-disagreed bits as follows:
\begin{itemize}
	 \item \textbf{All-agreed bits:} the subset of reliable code bits, whose decoded values are the same in \emph{all} previous decoding paths, and \emph{agrees} with the hard decision of the corresponding received symbol.
	 \item \textbf{All-disagreed bits:} the subset of reliable code bits, whose decoded values are the same in \emph{all} previous decoding paths, but \emph{disagrees} with the hard decision of the corresponding received symbol.
 \end{itemize}

Since both all-agreed bits and all-disagreed bits are reliable code bits, applying biased perturbation to both subsets will increase successful decoding probability in the next decoding attempts. However, this approach favors convergence over divergence, and thus cannot explore more codeword candidates in search for a correct one. Intuitively, as shown in Fig.~\ref{iSCL_example}, we propose the following $t$-attempt \textbf{adaptive biased perturbation} approach to strike a good balance between divergence and convergence.
\begin{enumerate}
	 \item In the 1st, 2nd, ... $(t-1)$-th attempts, the algorithm performs both exploitation and exploration by applying biased perturbation only to the all-disagreed bits.
	 \item In the $t$-th, that is, the last attempt, the algorithm maximizes exploitation by applying biased perturbation to both all-agreed bits and all-disagreed bits.
 \end{enumerate}

We formally describe the \textbf{biased perturbation} algorithm in Algorithm~\ref{algo:iterativenoise}.  It generates perturbation noise based on the received symbols $\mathbf{y}$, the perturbation noise power $\sigma_p^2$, and the set of all previously collected decoding paths $\mathcal{P}$. For each received symbol, the algorithm determines its hard decision $\bar{c_i}$ to be $1$ if $y_i<0$ and $0$ otherwise. It then compares $\bar{c_i}$ with the corresponding decoded bit $\hat{c_i}$ for all decoding paths in $\mathcal{P}$. If $\bar{c_i}\neq \hat{c_i}$ for all decoding paths in $\mathcal{P}$, the perturbation signal $p_i$ is calculated as $\frac{\sigma}{\sqrt{2}}(-1)^{\hat{x_i}}$. Otherwise, $p_i$ is sampled from a normal distribution $\mathcal{N}(0, \sigma^2)$.

\section{Performance evaluations}\label{sec:evaluation}
In this section, we present the simulation results of the SCL-perturbation decoding algorithm. We use CRC-aided polar codes with 16-bit CRC bits. The information set $\cal I$ is obtained by Gaussian approximation, where the constructing SNR is chosen to yield the best performance at BLER$=0.01$. As a benchmark, we implemented \emph{dynamic} SCL flip accoding to the \emph{state-of-the-art} method in~\cite{DynamicFlip}. Specifically, we designed the decoding parameter $\alpha$ for each SNR to get better performance (see Section~V.C in~\cite{DynamicFlip} for more details).

\subsection{Random perturbation}
In this subsection, we present the simulation results of the random SCL-perturbation decoding algorithm.

\subsubsection{Choosing perturbation power}
\begin{figure}[tbp]
	\centering
	\includegraphics[width=2.8in]{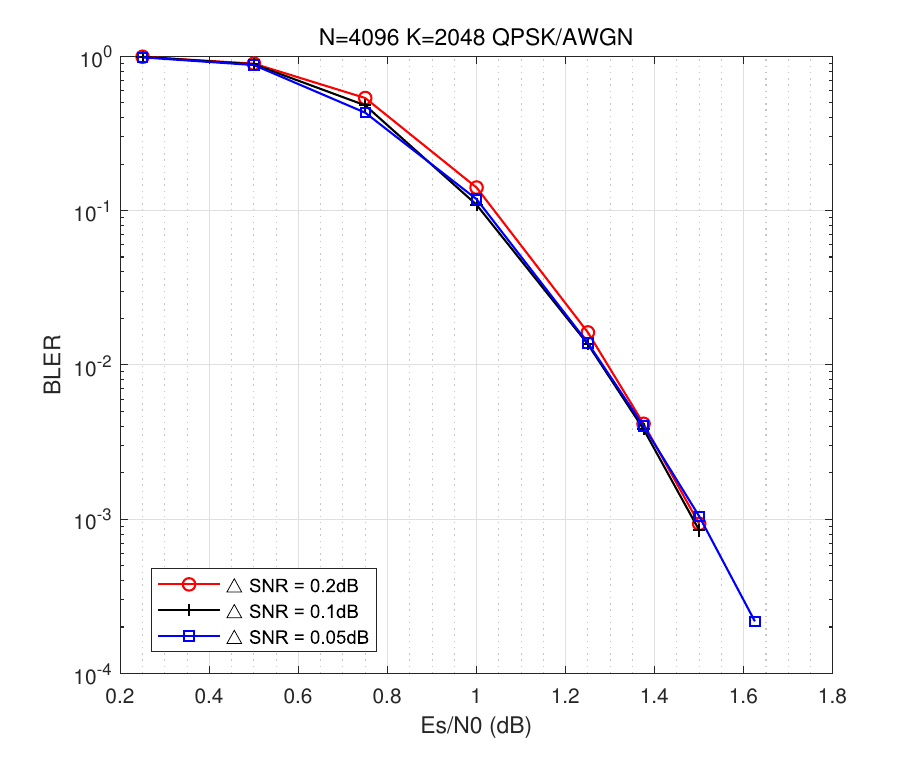}
	\caption{The optimal perturbation powers vary slightly at different SNRs.}
	\label{deltaSNR}
\end{figure}
In Fig.~\ref{deltaSNR}, we evaluate the decoding performance under different perturbation power levels. In this simulation, three perturbation powers $\Delta$SNR=0.2dB, 0.1dB and 0.05dB are used to decode $(N=4096, K=2048)$ polar codes using CA-SCL decoder with list size $L=8$, and $T=10$ maximum decoding attempts. It is observed that the optimal perturbation powers vary slightly at different SNRs.
At lower SNRs, a smaller perturbation power is preferred. In general, the performance is not very sensitive to the choice of $\Delta$SNR, making the SCL-perturbation algorithm robust for different scenarios.
For a balanced performance and the implementation simplicity, we choose a perturbation power of $\Delta$SNR=$0.1$dB for all code rates and lengths, and use it for all subsequent evaluations. This makes the proposed algorithm more hardware-friendly, while the parameters for SC/SCL-flip are basically case-dependent~\cite{DynamicFlip}.

\subsubsection{Different code lengths}
In Fig.\ref{simualtion}, we present the simulation results for coding rate $R = \frac{1}{2}$ and code lengths $N=\{256,1024,4096,16384\}$, and use SCL-8 as the component decoder. As seen, the SCL perturbation algorithm achieves performance comparable to a list-$16$ decoder, demonstrating the stability across different lengths.
On interesting observation is, compared to the SCL flip algorithm, the performance gain becomes more obvious as the code length increases. For $N=256$, SCL-flip provides slightly better performance. At $N=1024$, SCL-perturbation begins to outperform SCL-flip. For longer codes, e.g., at $N=16384$, we observe a clear advantage over SCL-flip.

\begin{figure}[tbp]
	\centering
	\includegraphics[width=2.8in]{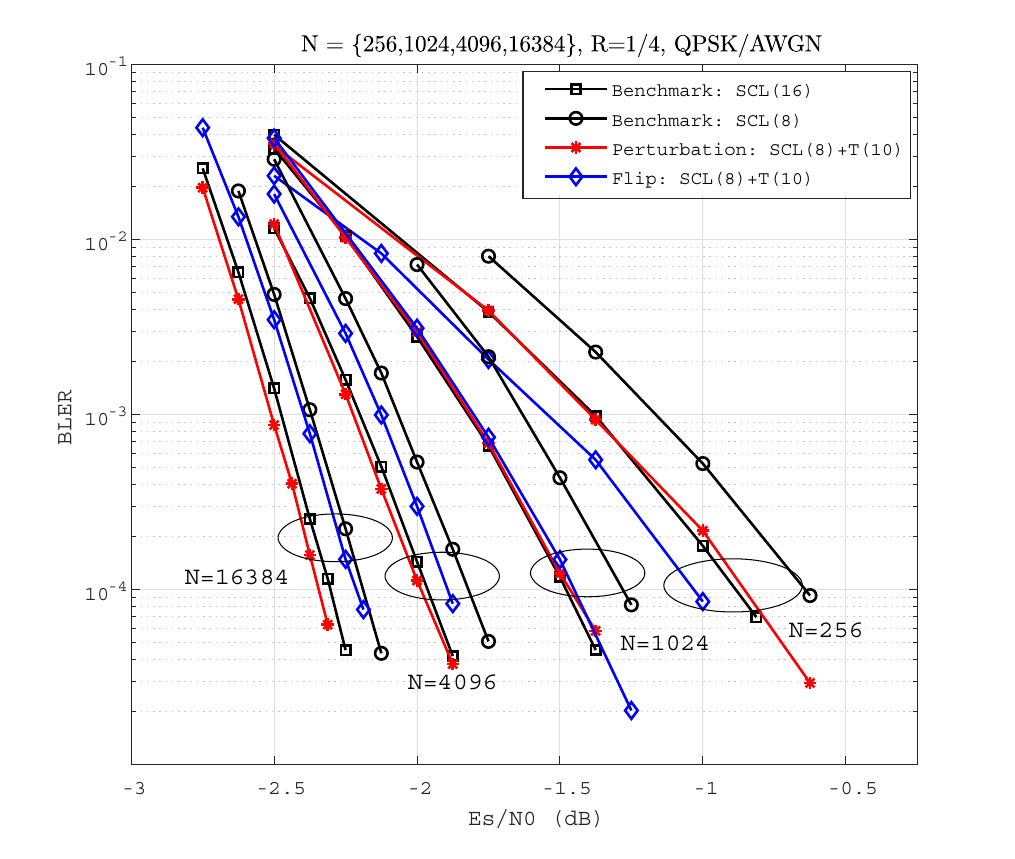}
	\caption{BLER performance for polar codes of $R=\frac{1}{4}$ and $N=\{256,1024,4096,16384\}$.}
	\label{sim025}
\end{figure}

\begin{figure}[tbp]
	\centering
	\includegraphics[width=2.8in]{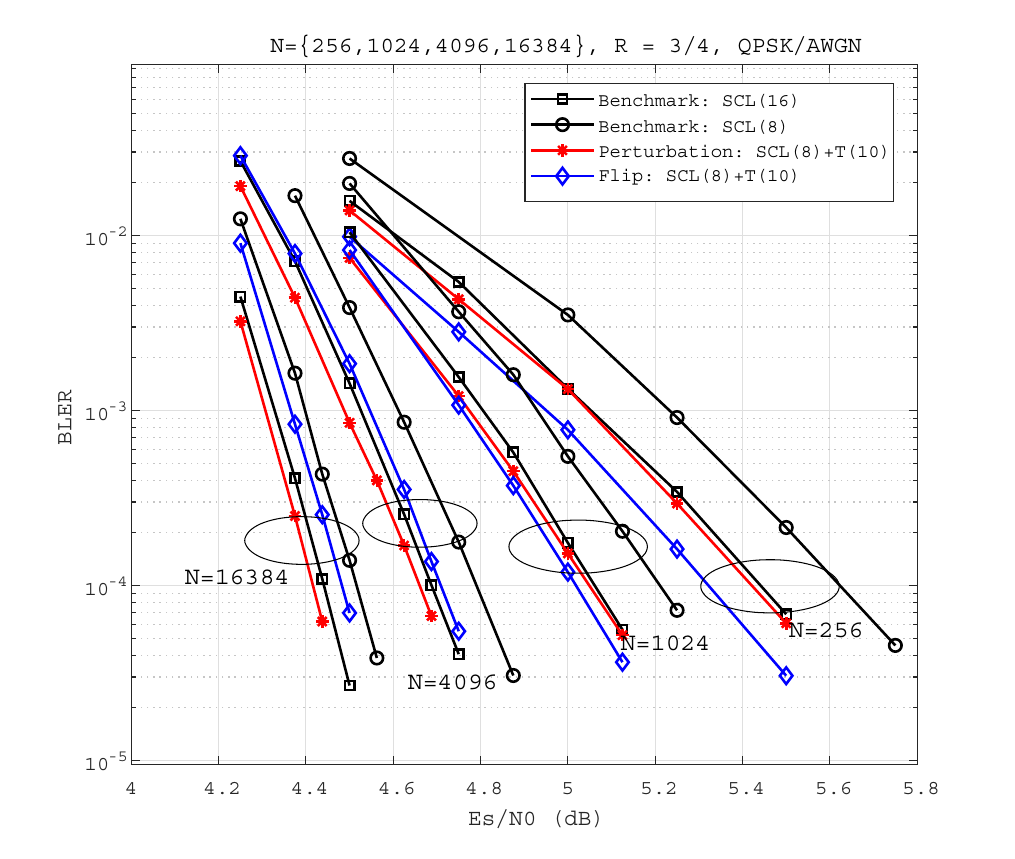}
	\caption{BLER performance for polar codes of $R=\frac{3}{4}$ and $N=\{256,1024,4096,16384\}$.}
	\label{sim075}
\end{figure}

In Fig.~\ref{sim025} to Fig.~\ref{sim075}, we also present the simulation results for code rate $R = \{\frac{1}{4}, \frac{3}{4}\}$ to show SCL-perturbation's stable gain.

\subsubsection{Different code rates}
\begin{figure}[tbp]
	\centering
	\includegraphics[width=2.8in]{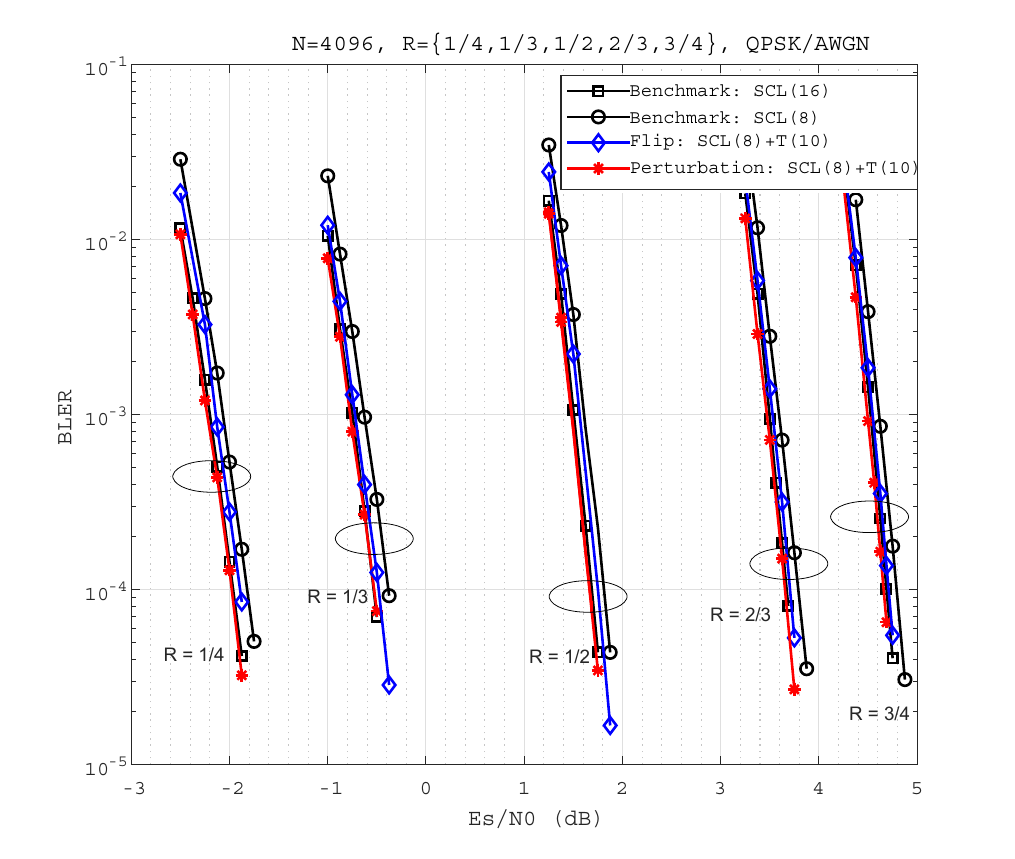}
	\caption{BLER performance at different coding rates.}
	\label{rates}
\end{figure}

Then, we compare the performance gain across different coding rates. The simulation results for $N=4096$ and $R = \{\frac{1}{4}, \frac{1}{3}, \frac{1}{2}, \frac{2}{3}, \frac{3}{4}\}$ are presented in Fig.~\ref{rates}.
It is observed that SCL-perturbation achieves performance comparable to a list-$16$ decoder for all code rates.
At lower rates, the gain is larger. For rate-$1/4$, the gain is $0.1$dB over SCL-flip.

\subsubsection{Different list sizes}
In this subsection, we compare the perturbation gain and flipping gain using component SCL decoders with  different list sizes.
In particular, we fix $N=4096$ and $R=1/2$, and try list sizes $4$, $8$, and $16$.
Fig.~\ref{fig:variouslist} shows that list size does impact the performance gain. 
For an SCL-4 decoder, 10 decoding attempts yield $0.18$dB perturbation gain, but only $0.12$dB flipping gain. The latter is $33\%$ less gain in dB.
But when the component SCL decoder's list size is $16$, the performance gains using perturbation and flipping are $0.1$dB and $0.02$dB, respectively. In other words, flipping does not work well for SCL decoder.
\begin{figure}[tbp]
	\centering
	\includegraphics[width=3.1in]{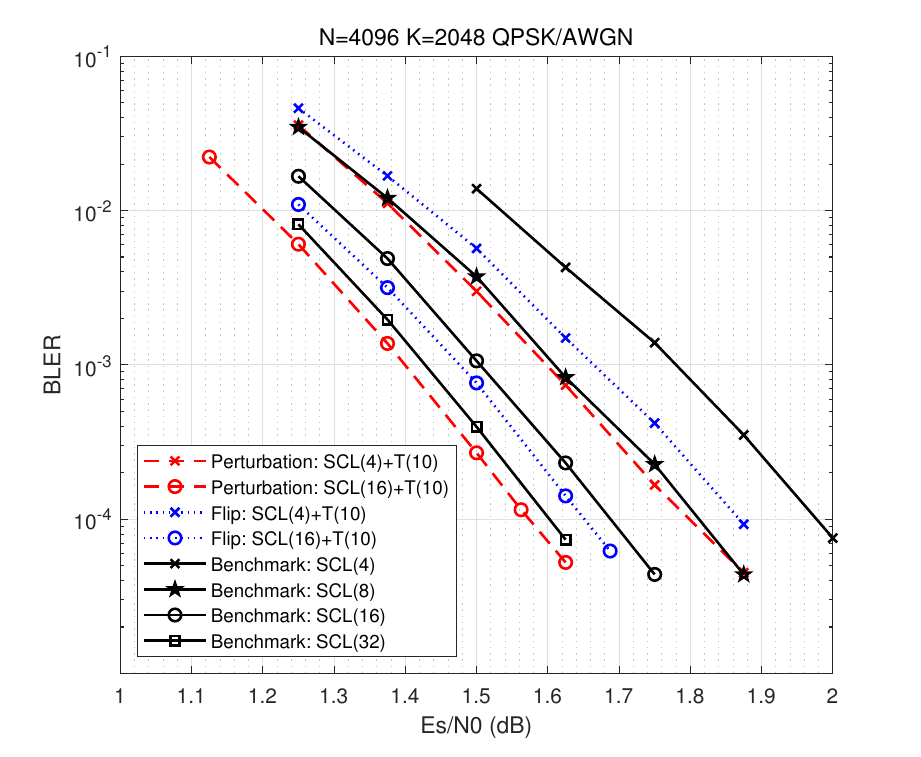}
	\caption{BLER performance with different list sizes.}
	\label{fig:variouslist}
\end{figure}

\subsubsection{Different numbers of decoding attempts}
\begin{figure}[tbp]
	\centering
	\includegraphics[width=2.8in]{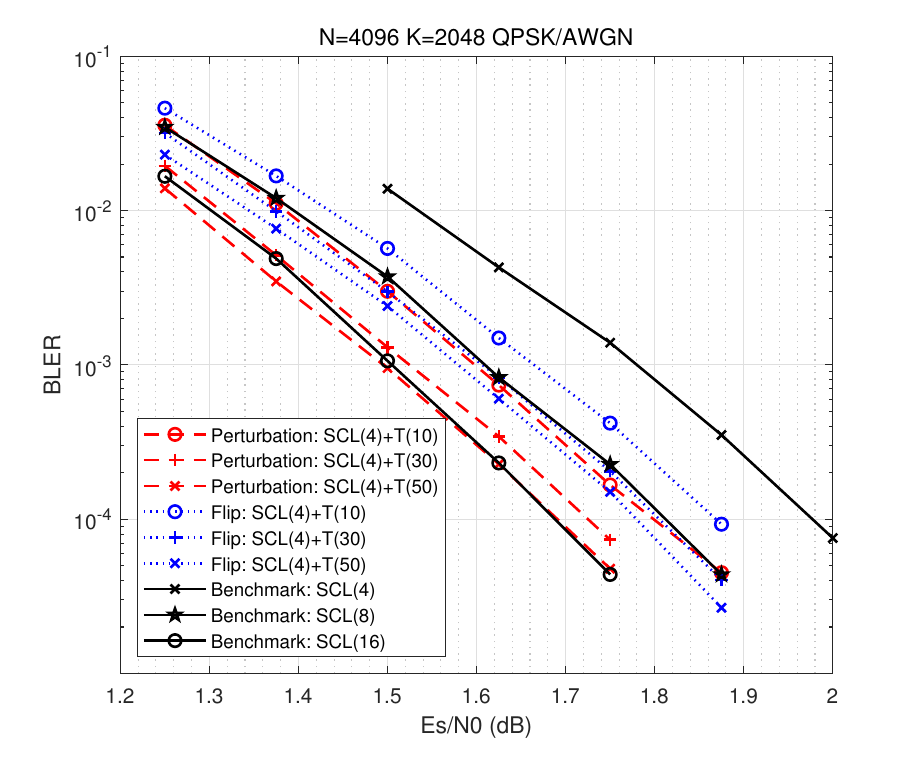}
	\caption{BLER performance under different number of decoding attempts.}
	\label{fig:variousnumber}
\end{figure}

We investigate how much additional gain can SCL-perturbation and SCL-flip obtain as the number of decoding attempts increase.
We tested $T = \{10, 30, 50\}$ for $N=4096, K=2048$, as shown in Fig.\ref{fig:variousnumber}.

SCL-perturbation turns out to be more efficient than SCL-flip, because 10 rounds of perturbation outperform 30 rounds of the flipping. When both algorithms increase their attempts from $T=10$ to $30$, the additional gains at $BLER=10^{-3}$, are 0.08dB and 0.05dB, respectively. This shows that the perturbation benefits more from additional decoding attempts, while more flipping does not help much.

\subsubsection{Improved slope of BLER curves}
\begin{figure}[tbp]
	\centering
	\includegraphics[width=2.8in]{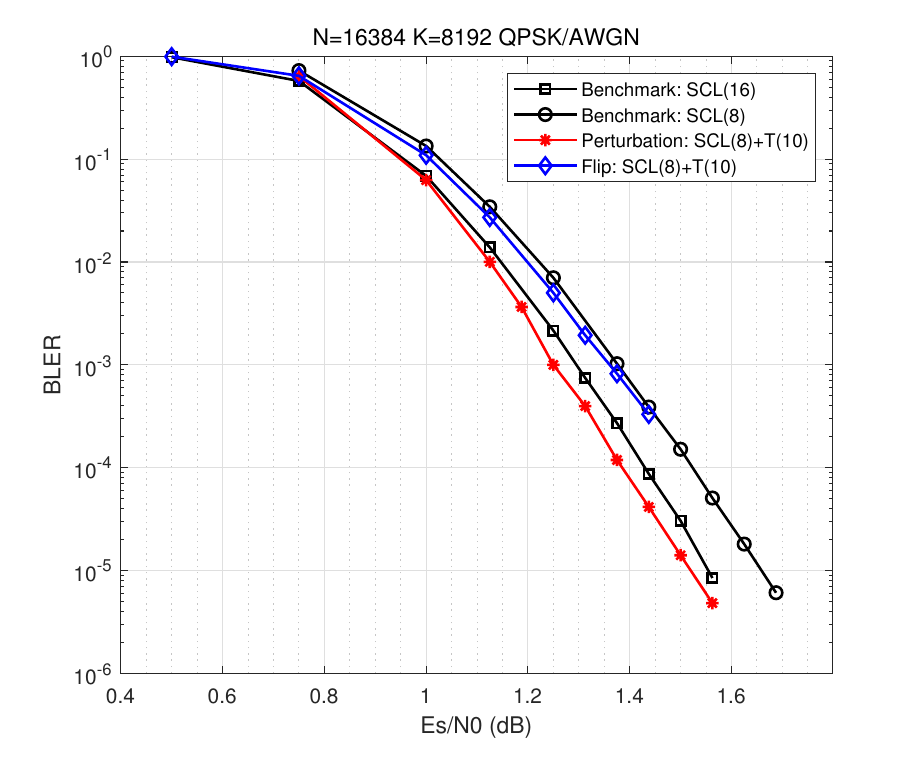}
	\caption{SCL-perturbation leads to a steeper slope in the BLER curves, and thus the gain widens at lower BLER region.}
	\label{fig:slope}
\end{figure}

As shown in Fig.~\ref{fig:slope}, SCL-perturbation with component SCL-8 decoder outperforms SCL-16 at $N=16384$ and $R=\frac{1}{2}$, while the flipping gain vanishes.
One remarkable observation is that SCL-perturbation leads to a steeper slope in the BLER curves, and thus the gain widens at lower BLER region. Note that the slope is even steeper than that of SCL-16.

\subsection{Biased perturbation}
\label{performance_bp}
This subsection presents the simulation results of the biased SCL-perturbation decoding algorithm.

\subsubsection{Different biased perturbation methods}
We evaluate the proposed adaptive biased perturbation scheme. As discussed in Section~\ref{BP}, this method applies different biased perturbation methods over multiple attempts. Specifically, it adopts \textbf{partly-biased perturbation} in the first $t-1$ attempts, and switch to \textbf{all-biased perturbation} in the final attempt. For partly-biased perturbation, biased perturbations are applied only to the all-disagreed bits. For all-biased perturbation, biased perturbations are applied to both all-agreed and all-disagreed bits. For comparison, we evaluate all three biased perturbation schemes, as follows.

\begin{itemize}
	\item \textbf{Partly-biased perturbation}: always apply biased perturbation only to the all-disagreed bits. This scheme favors divergence and exploration.
	\item \textbf{All-biased perturbation}: always apply biased perturbation to both all-agreed and all-disagreed bits. This scheme favors convergence and exploitation.
	\item \textbf{Adaptive biased perturbation}: apply \textbf{partly-biased perturbation} in the first $t-1$ attempts, and switch to \textbf{all-biased perturbation} in the final $t$-th attempt. This scheme seeks a balance between exploration and exploitation.
\end{itemize}

Aiming at practical hardware implementations, we limit our evaluations to low-complexity scenarios. The widely implemented SCL-$8$ is chosen as the component decoder and the maximum decoding attempts are limited to $T_{\max}=\{2, 3, 5\}$. We also include SCL and random SCL-perturbation with list size $L=8$ decoders as benchmarks.

\begin{figure}[tbp]
	\centering
	\includegraphics[width=3in]{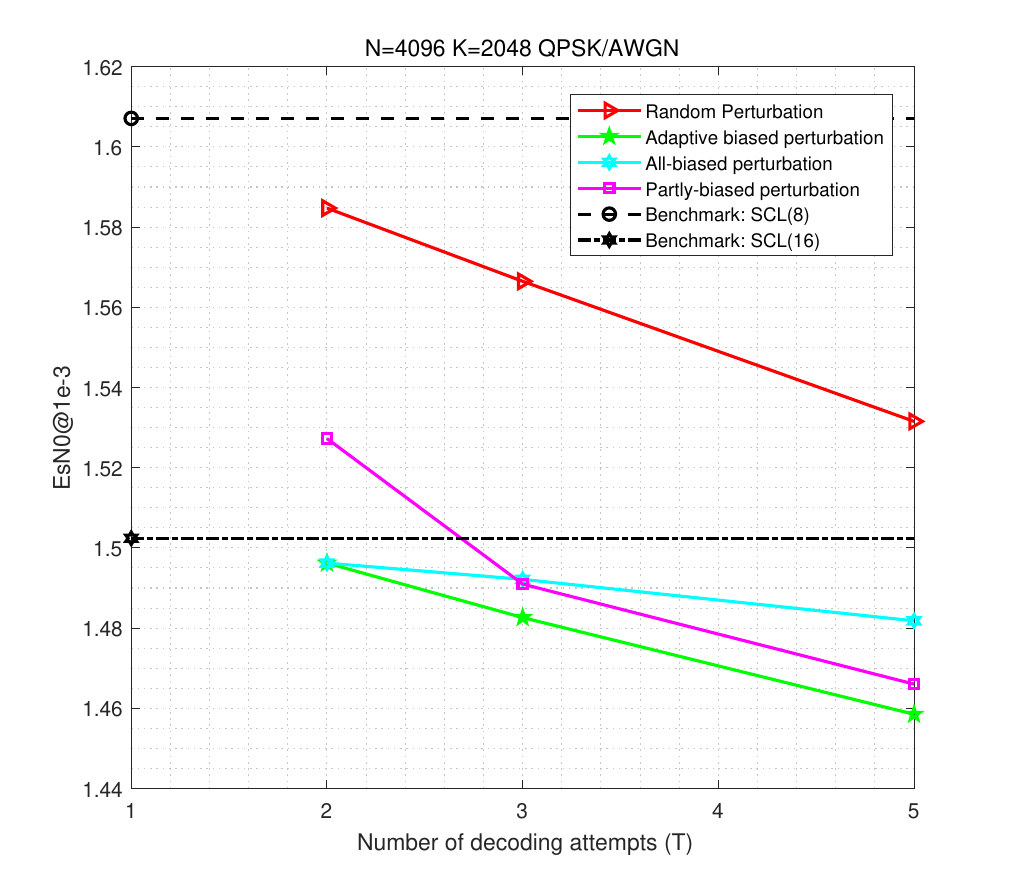}
	\caption{The proposed adaptive biased approach outperforms both the partly-biased perturbation and all-biased perturbation schemes.}
	\label{choosbiasedbits}
\end{figure}

The evaluation results for $(N=4096, K=2048)$ polar codes are presented in Fig.~\ref{choosbiasedbits}. Firstly, all the biased perturbation schemes are more efficient than the random perturbation scheme. Specifically, all the biased perturbation schemes can achieve the effect of doubling list size with merely $T=2 \backsim 3$ decoding attempts. Meanwhile, it takes random perturbation more than $T=5$ attempts to achieve the same performance. This demonstrates the effectiveness of the biased-perturbation algorithm.

Secondly, the proposed adaptive biased approach outperforms both the partly-biased perturbation and all-biased perturbation schemes. This confirms the necessity to strike a balance between divergence and convergence.

Finally, we compare the two non-adaptive biased perturbation schemes. With a maximum of $2$ decoding attempts, \textbf{all-biased perturbation} provides better performance than \textbf{partly-biased perturbation}. However, as the number of decoding attempts increases, the results are reversed. More attempts with all-biased perturbation do not provide much additional gain, with $T=5$ and $T=2$ achieving nearly the same performance. In contrast, for partly-biased perturbation, $T=5$ brings a $0.06$~dB gain over $T=2$. This indicates the need for a divergent decoder in the long run, through collecting more new candidate decoding paths in the first a few attempts. If the decoder always converges toward a particular direction, it may end up being stuck with certain incorrect codewords.

In the following, we adopt the adaptive biased perturbation scheme in all simulations due to its best performance and efficiency.

\subsubsection{Different code lengths}
In Fig.~\ref{iSCL_performance_05}, we present the simulation results for coding rate $R = \frac{1}{2}$ and code lengths $N=\{256, 1024, 4096, 16384\}$, using SCL-8 as the component decoder. To ensure low implementation complexity, we limit the maximum number of decoding attempts to only 2.

As shown, in the long code length regime, biased perturbation provides significantly better performance than random perturbation. Specifically, for $N=16384$, biased perturbation achieves a 0.125 dB gain, whereas random perturbation provides less than a 0.03 dB gain. To achieve the performance of SCL-16, the biased perturbation algorithm requires at most two decoding attempts, which greatly reduces the complexity compared to the random perturbation scheme.

\begin{figure}[tbp]
	\centering
	\includegraphics[width=3in]{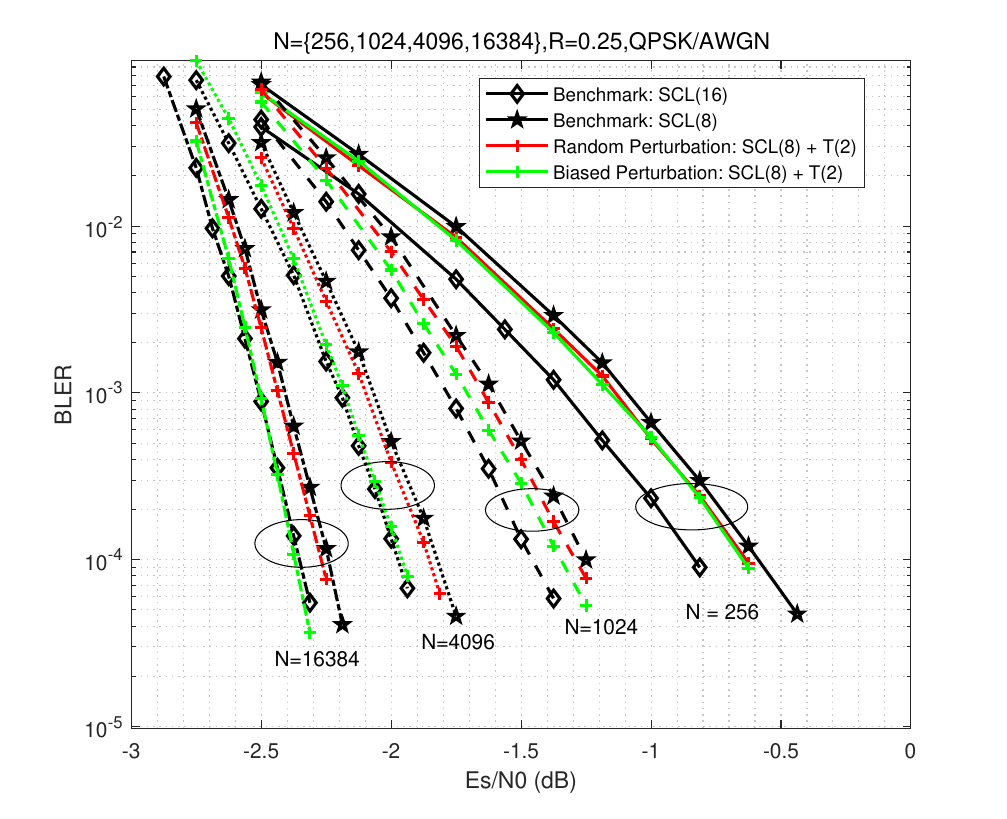}
	\caption{BLER performance for polar codes of $R=0.25$ and $N=\{256,1024,4096,16384\}$.}
	\label{biassim025}
\end{figure}

\begin{figure}[tbp]
	\centering
	\includegraphics[width=3in]{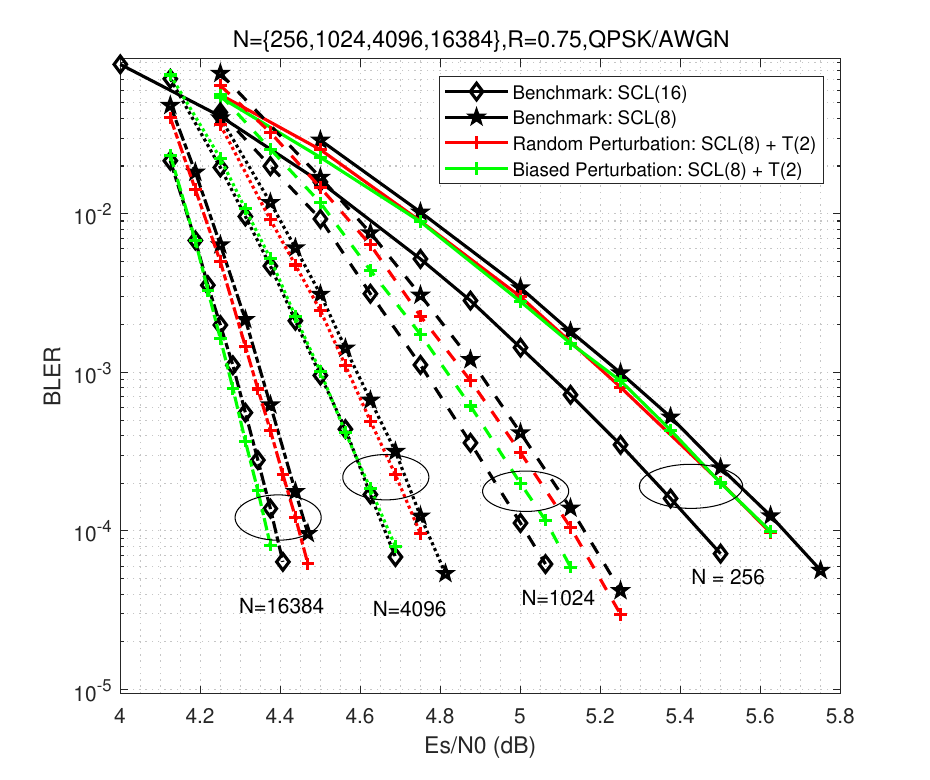}
	\caption{BLER performance for polar codes of $R=0.75$ and $N=\{256,1024,4096,16384\}$.}
	\label{biassim075}
\end{figure}

In Figs.~\ref{biassim025} to \ref{biassim075}, we additionally include the simulation results for coding rates $R = \{\frac{1}{4}, \frac{3}{4}\}$ to show the stable gain provided by biased perturbation across different code rates.

\subsubsection{Different list sizes}
We investigate the biased perturbation algorithm with component decoders of different list sizes.

In the simulation, we use SCL component decoders ($L = 4, 8, 16, 32$) to decode a $(N=4096, K=2048)$ polar code. As benchmarks, we compare with the results of the SCL and random perturbation schemes.

As shown in Fig.~\ref{fig:variouslist}, the biased perturbation algorithm provides stable performance gain over the random perturbation algorithm.

To achieve double-list-size SCL performance, the required number of decoding attempts decreases as the list size increases. Specifically, with an SCL-4 component decoder, one extra decoding attempt using biased perturbation yields performance slightly worse than that of an SCL-8 decoder. With an SCL-8 component decoder, one extra decoding attempt using biased perturbation achieves the same performance as an SCL-16 decoder. Using SCL-16 or SCL-32 as component decoders, one extra decoding attempt using biased perturbation provides better performance than the respective double-list-size SCL decoder.
\begin{figure}[tbp]
	\centering
	\includegraphics[width=3.1in]{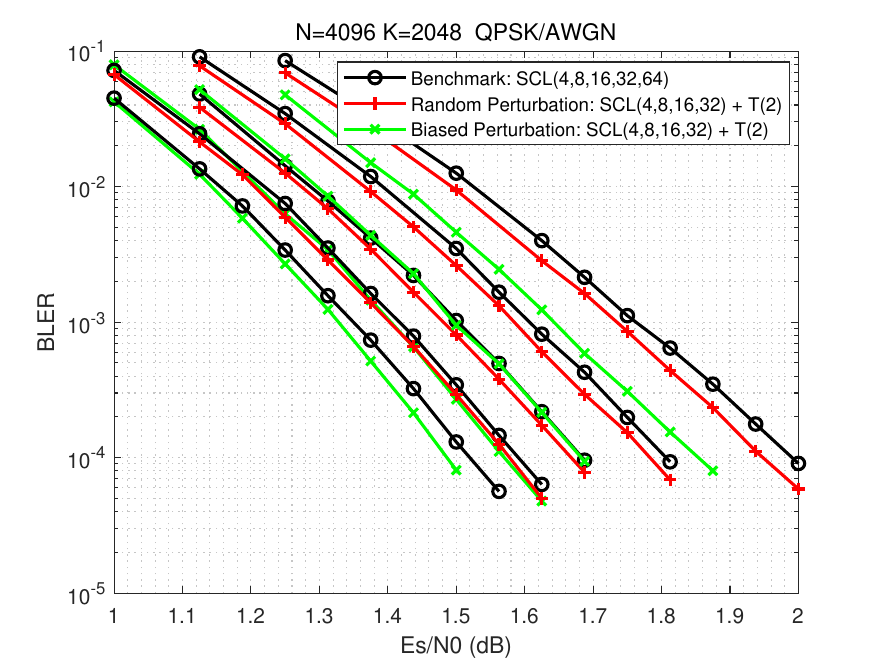}
	\caption{BLER performance with component decoders of different list sizes.}
	\label{fig:variouslist}
\end{figure}

The above thorough simulation results demonstrate a highly-efficient SCL-based decoder. It far outperforms existing algorithms in the low-complexity and large block length regime. For hardware implementation, this biased perturbation approach offers an alternative to increasing the list size. Thus it has a particular advantage of not requiring additional chip area.

\section{Complexity evaluation}\label{sec:complexity}
In this section, we evaluate decoding complexity with three metrics: worst-case computational complexity, average computational complexity, and hardware implementation complexity. We also compare SCL-perturbation with both SCL-flip and SCL.

\subsection{Worst-case computational complexity}
For an SCL decoder with list size $L$, the complexity is in the order of $\mathcal{O}(LN\log(N))$~\cite{TalVardy}. For SCL-perturbation, the complexity is $\mathcal{O}(TLN\log(N))$, i.e., $T$ times that of SCL.

For random perturbation, the worst-case complexity is about $5$ times that of a double-list-size SCL decoder, since simulation results show that $T=10$ is required to achieve a double-list-size gain.

For the enhancement with biased perturbation, the worst-case complexity is the same as that of a double-list-size SCL decoder. As shown in Section~\ref{performance_bp}, $T=2$ is sufficient to achieve a double-list-size gain for long polar codes.

Compared to SCL-flip, the proposed SCL-perturbation scheme requires much lower worst-case complexity to achieve a double-list-size gain. For $N=4096, K=2048$ polar codes, SCL-flip requires $T=30$ decoding attempts (see Fig.~\ref{fig:variousnumber}), whereas SCL-perturbation requires only $T=2$. In other words, SCL-flip requires $15$ times the worst-case complexity of SCL-perturbation. For larger code lengths, the SCL-flip-to-SCL-perturbation complexity ratio is much higher, making SCL-flip unsuitable for long polar codes.

\subsection{Average computational complexity}
The advantage of SCL-perturbation is also exhibited through average computational complexity. Once we obtain a decoding candidate that passes CRC check, we early terminate the decoding and record the number of decoding attempts used. In practice, the decoder circuit can be switched off to save energy.

\begin{figure}[tbp]
	\centering
	\includegraphics[width=3in]{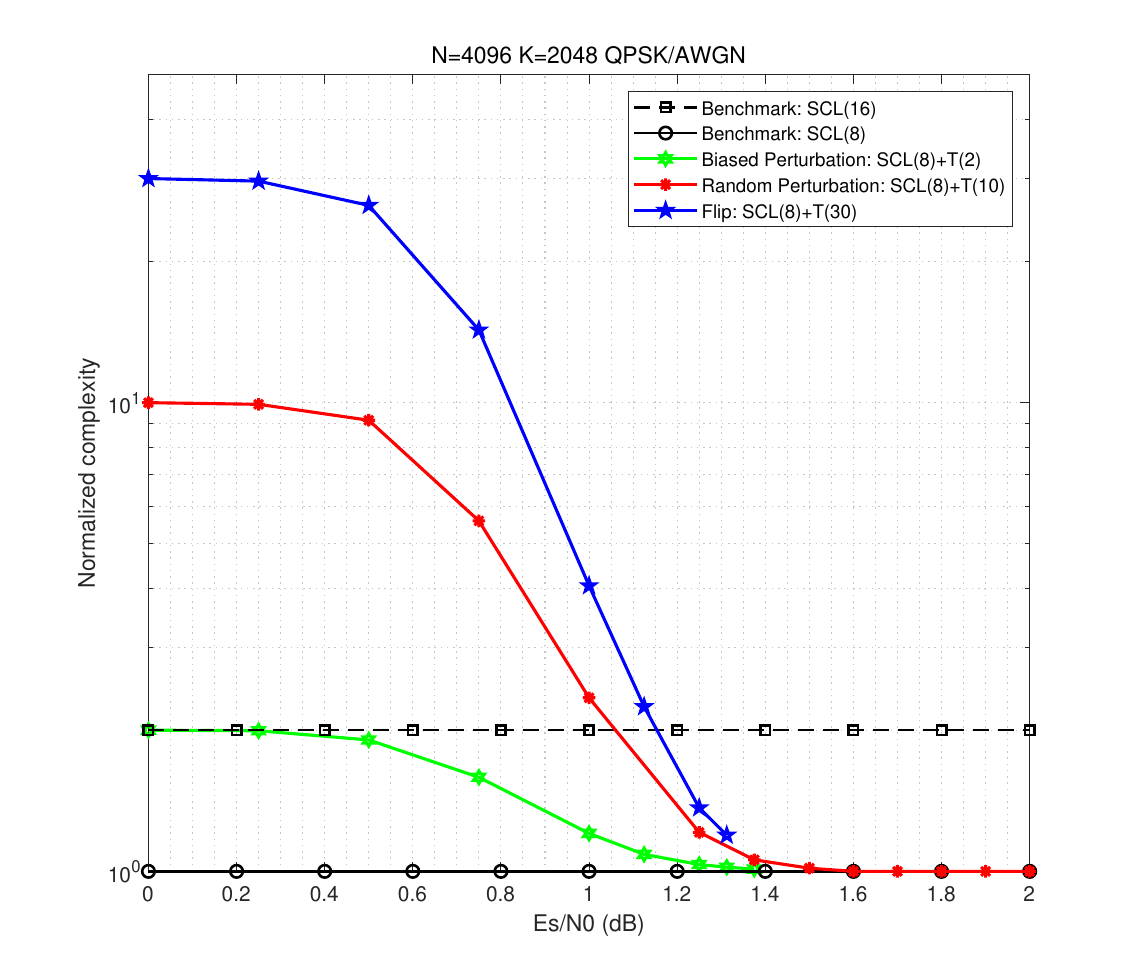}
	\caption{Average complexity of SCL perturbation normalized with respect to the complexity of the component decoder. The maximum number of decoding attempts is set to achieve double-list performance.}
	\label{average}
\end{figure}
In Fig.~\ref{average}, we compare the average computational complexity of SCL, SCL-perturbation, and SCL-flip at different SNRs. At low SNR, both SCL-perturbation and SCL-flip exhibit $T$ times of SCL complexity. As SNR increases, the average complexity of both SCL-perturbation and SCL-flip drops to SCL complexity, which is expected. 

\subsection{Hardware implementation complexity}
\begin{figure}[tbp]
	\centering
	\includegraphics[width=2.3in]{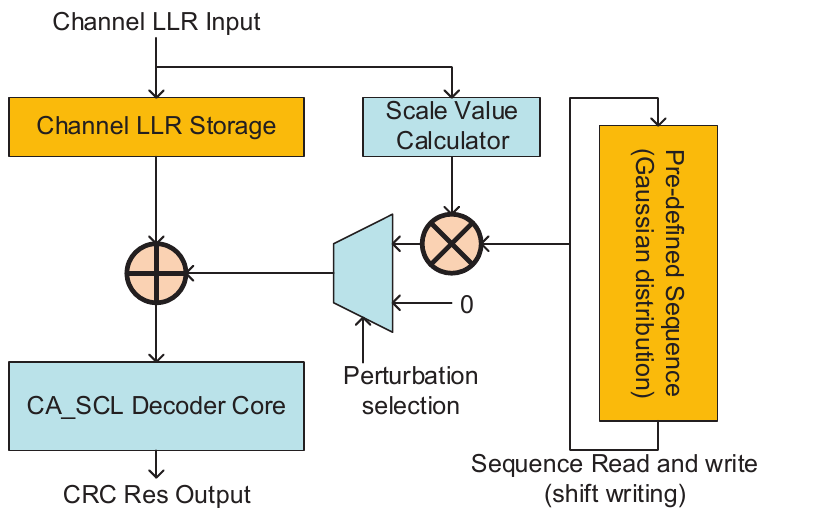}
	\caption{Decoding architecture of SCL perturbation algorithm.}
	\label{architecture}
\end{figure}

The advantages of SCL-perturbation are two-fold:
\begin{itemize}
  \item The performance gain is achieved without increasing the list size. In hardware, this means no additional circuit. Note that a double-list-decoder not only doubles the memory size and processing elements, but requires additional sorting network to handle the larger list size. In this sense, SCL-perturbation is a hardware-friendly decoding algorithm.
  \item Compared to SCL-flip, the perturbation operations are also much simpler. As shown in the decoding architecture in Fig.~\ref{architecture}, it only requires a module to add pseudo-random noise to the received symbols, which involves a few additional shift registers and can be implemented in parallel. However, the flipping operations require pre-calculations of certain reliability metrics and sorting them in order to identify the bit-flip positions. As we know, a sorter is quite expensive in hardware, and when the code length increases, the metrics to be sorted increase too, making the sorter network prohibitively large, if not infeasible. Therefore, flipping operations require much more computational and memory resources than perturbation operations in hardware.
\end{itemize}

\section{Deep learning based perturbation}\label{sec:discussion}
In this section, we discuss the possible deep learning based approach to further improve decoding performance and efficiency.

In section~\ref{BP}, the proposed biased perturbation approach is heuristic in several ways. First, the perturbation strength for both random perturbation and biased perturbation is empirically determined. Second, the subset of code bits for biased perturbation for each decoding attempt is also heuristically determined.

A theoretical modeling of the SCL-perturbation decoder and the corresponding optimal perturbation strategy are still open problems. That is to say, although the proposed biased perturbation scheme exhibits excellent performance and efficiency, we still do not know whether it is optimal and how much room for further improvement. After all, a theoretical analysis of the component SCL decoder is still missing~\cite{SCLanalysis}, in particular the path sorting, splitting and pruning processes.

When a rigorous theory can not be established, deep learning based approaches usually offer satisfying alternatives. In~\cite{AICoding}, deep learning, reinforcement learning, and genetic algorithms are adopted for optimizing code constructions. In~\cite{LearningFlip}, a long short-term memory (LSTM) recurrent neural network (RNN) is employed to select the bit-flip positions to enhance an SCL-flip decoder.

\begin{figure}[tbp]
	\centering
	\includegraphics[width=2.5in]{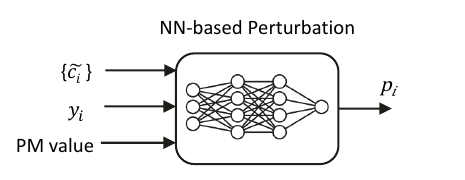}
	\caption{Neural network based perturbation value generators.}
	\label{ai_model}
\end{figure}
When applied to the perturbation algorithm, a deep neural network (NN) can be employed to select (i) perturbation strength, (ii) bit positions for perturbation, and (iii) different perturbation methods, e.g., random or biased perturbation.

In our initial attempt, we used a neural network to directly determine the perturbation values. Specifically, we use a fully connected network to act as the perturbation generator. As shown in Fig.~\ref{ai_model}, it takes the received symbols and decoding results from previous attempts as inputs. Since biased perturbation is also based on these factors, this ensures that the network has sufficient information to \emph{learn} the biased perturbation. To further enhance accuracy, we additionally feed the path metric (PM) values of each decoding attempt to the network.

Supervised learning is used to train the model. A reward function is defined according to BLER performance~\cite{AICoding}, and the neural network is trained via backpropagation~\cite{Backpropagation}. The neural network model can be easily trained by an open-source platform for machine learning, such as TensorFlow~\cite{TensorFlow}.

\begin{figure}[tbp]
	\centering
	\includegraphics[width=3in]{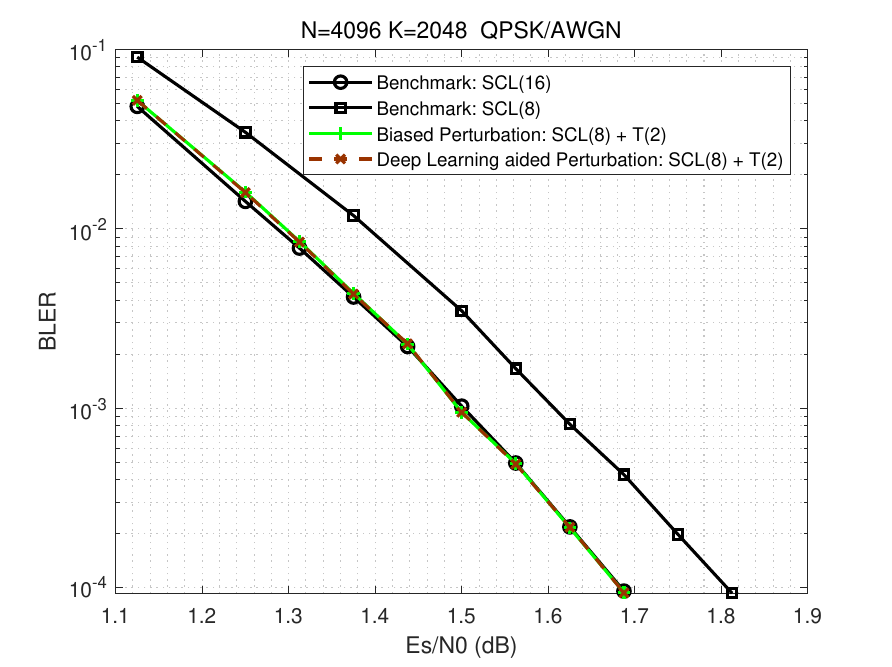}
	\caption{Deep learning-aided perturbation achieves nearly the same performance as biased perturbation.}
	\label{fig:deeplearning_aided}
\end{figure}

To evaluate the performance, we applied the well-trained network to the SCL-perturbation algorithms. The simulation results in Fig.~\ref{fig:deeplearning_aided} show that the deep learning-aided approach achieves nearly the same performance as the biased SCL perturbation approaches. This indicates the great potential of deep learning in further improving SCL-perturbation algorithms.

\section{Conclusion}\label{sec:conclusion}
In this work, we study perturbation as an enhancement technique to improve the SCL decoding performance. The most striking observation is that, with the same number of decoding attempts, the performance gain of SCL-perturbation is higher than SCL-flip at larger code length. We also found that the performance and efficiency of SCL-perturbation can be significantly improved with an adaptive biased perturbation scheme. The best result so far is that with only one extra decoding attempt (two in total), SCL-perturbation can achieve the performance of a double-list-size SCL decoder. The observation is consolidated through extensive simulations under various code rates, code lengths, list sizes and numbers of decoding attempts. In particular, the performance gain does not vanish as code length increases, and its demonstrated efficiency far outperforms other SCL enhancements such as SCL-flip and SCL-AE. The advantages of SCL-perturbation are also demonstrated by its lower computational and hardware complexity.

\section{Acknowledgement}
The authors thank Dr. Kangjian Qin and Dr. Zhicheng Liu for the fruitful discussions that inspired this work.

\end{document}